\documentclass[aip,jcp,preprint,superscriptaddress,amsmath]{revtex4-1}
\usepackage{graphicx}
\newcommand{\avg}[1]{\left\langle #1 \right\rangle}
\begin{document}
\title{Intensity interferometry of single x-ray pulses from a synchrotron storage ring}
\author{A. Singer}
\altaffiliation{Present address: University of California, San Diego, La Jolla, California 92093, USA}
\affiliation{Deutsches Elektronen-Synchrotron DESY, Notkestra\ss{}e 85, D-22607 Hamburg, Germany}
\author{U. Lorenz}
\altaffiliation{Present address: Department of Chemistry, University of Potsdam, Karl-Liebknecht-Strasse 24-25, D-14476 Potsdam OT Golm}
\affiliation{Deutsches Elektronen-Synchrotron DESY, Notkestra\ss{}e 85, D-22607 Hamburg, Germany}
\author{A. Marras}
\author{A. Klyuev}
\author{J. Becker}
\affiliation{Center for Free-Electron Lasers, Notkestrasse 85, D-22607 Hamburg, Germany}
\author{K. Schlage}
\affiliation{Deutsches Elektronen-Synchrotron DESY, Notkestra\ss{}e 85, D-22607 Hamburg, Germany}
\author{P. Skopintsev}
\affiliation{Deutsches Elektronen-Synchrotron DESY, Notkestra\ss{}e 85, D-22607 Hamburg, Germany}
\affiliation{National Research Center 'Kurchatov Institute', Kurchatov square 1, 123182 Moscow, Russia}
\author{O. Gorobtsov}
\affiliation{Deutsches Elektronen-Synchrotron DESY, Notkestra\ss{}e 85, D-22607 Hamburg, Germany}
\affiliation{National Research Center 'Kurchatov Institute', Kurchatov square 1, 123182 Moscow, Russia}
\author{A. Shabalin}
\author{H.-C. Wille}
\author{H. Franz}
\affiliation{Deutsches Elektronen-Synchrotron DESY, Notkestra\ss{}e 85, D-22607 Hamburg, Germany}
\author{H. Graafsma}
\affiliation{Center for Free-Electron Lasers, Notkestrasse 85, D-22607 Hamburg, Germany}
\affiliation{Mid Sweden University, S-851 70 Sundsvall, Sweden}
\author{I. A. Vartanyants}
\altaffiliation[Corresponding author:]{Ivan.Vartaniants@desy.de}
\affiliation{Deutsches Elektronen-Synchrotron DESY, Notkestra\ss{}e 85, D-22607 Hamburg, Germany}
\affiliation{National Research Nuclear University, ''MEPhI'', 115409 Moscow, Russia}

\date{\today}

\begin{abstract}
We report on measurements of second-order intensity correlations at the high brilliance storage ring PETRA III
using a prototype of the newly developed Adaptive Gain Integrating Pixel Detector (AGIPD).
The detector recorded individual synchrotron radiation pulses with an x-ray photon energy of 14.4 keV and repetition rate of about 5 MHz.
The second-order intensity correlation function was measured simultaneously at different spatial separations that allowed to determine the transverse coherence length at these x-ray energies. The measured values are in a good agreement with theoretical simulations based on the Gaussian Schell-model.
\end{abstract}
\maketitle

Third generation synchrotrons are nowadays the principal sources of high-brilliance x-ray radiation.
They generate beams with a high coherent flux, which are particularly useful for newly developed imaging techniques, such as coherent diffraction imaging (CDI) \cite{ChapmanNatPhot2010,Vartanyants2010,MancusoJB2010,VartanyantsBook} and ptychography \cite{RodenburgPRL2007,ThibaultScience2008}, as well as for studying dynamics by x-ray photon correlation spectroscopy \cite{GruebelJAC2004}.
Measurements of the transverse coherence are vital for the success of these coherence based applications.
Such measurements can also be used to monitor the size of the x-ray source, which can yield information on the electron bunch trajectory in the undulator and help to optimize the source parameters.

Different techniques can be applied to measure the transverse coherence in the x-ray range.
For soft x-rays, Young's double pinhole experiment was successfully used at synchrotron and free-electron laser sources \cite{ChangOptCommun2000,PatersonOptCommun2001,VartanyantsPRL2011,SingerOptExp2012}.
Unfortunately, this approach can not be directly extended to hard x-rays due to their high penetration depth.
In the hard x-ray range different approaches to determine the transverse coherence were implemented such as scattering from a thin wire \cite{KohnPRL2000}, shearing interferometry \cite{PfeifferPRL2005}, scattering on colloidal samples \cite{AlaimoPRL2009,GuttPRL2012}, or bi-lenses \cite{SnigirevPRL2009}.
In all these methods, amplitudes are correlated, and the coherence is measured through the visibility of interference fringes.
Alternatively, correlations between intensities can be measured to obtain the same information about the source.
Intensity correlation methods are particularly attractive for two reasons: phase fluctuations due to optics vibrations etc. are mitigated because no phases are measured, and no scattering sample is required, which might introduce uncertainties due to imperfections.

It is well established nowadays that coherence properties of a thermal (chaotic) source can be determined through coincident detection of photons, as first realized by Hanbury Brown and Twiss (HBT) \cite{BrownNat1956,BrownNat1956_}.
The key quantity in such an experiment is the normalized second-order intensity correlation function \cite{MW1995}
\begin{equation}
	g^{(2)}(\mathbf r_1,\mathbf r_2)=\frac{\avg{I(\mathbf r_1) I(\mathbf r_2)}}{\avg{I(\mathbf r_1)}\avg{I(\mathbf r_2)}},
\label{eq:2nd_corr}
\end{equation}
where $I(\mathbf r)$ is the intensity measured at position $\mathbf r$ and $\avg{\cdots}$ denotes the average over a large ensemble of independent measurements.
It can be shown that the intensity correlation function originating from pulsed chaotic sources can be described in terms of the amplitude correlation function, also known as the complex coherence function $\gamma(\mathbf r_1,\mathbf r_2)$ as \cite{IkonenPRL1992}
\begin{equation}
	g^{(2)}(\mathbf r_1,\mathbf r_2)=1+\zeta\cdot\left\vert\gamma(\mathbf r_1,\mathbf r_2)\right\vert^2,
\label{eq:chaotic}
\end{equation}
where $\zeta=1/M_l$ is the contrast value, $M_l=T/\tau_c$ is the number of longitudinal modes of radiation field, $T$ is the pulse duration, and $\tau_c$ is the coherence time, which scales inversely with the bandwidth and can be tuned by a monochromator.
For $\mathbf{r}_1 = \mathbf{r}_2$, eq.~\eqref{eq:chaotic} contains an additional term $1/\overline{N}$, where $\overline{N}$ is the average number of detected photons \cite{G1985} (see also Supplementary Material).
This last term is particularly important for low photon statistics.
The modulus of the complex coherence function $|\gamma(\mathbf r_1,\mathbf r_2)|$ is a measure of the visibility of interference fringes in a Young's double pinhole experiment \cite{G1985,MW1995}.

There has been considerable interest in performing HBT experiments at synchrotron sources \cite{KunimuneJSR1997,GluskinJSynchRad1999,YabashiPRL2001,YabashiPRL2002}.
The key to the success of these experiments was the development of high resolution monochromators and the use of avalanche photodiodes (APD), which have sufficient temporal resolution to discriminate single synchrotron radiation pulses.
However, a single measurement of such type gives the intensity correlation function $g^{(2)}(\mathbf{r}_1,\mathbf{r}_2)$ only for one pair of transverse coordinates $(\mathbf{r}_1$ and $\mathbf{r}_2)$; to map out the full transverse correlation function, the measurement has to be repeated multiple times.
Here we propose to use the new Adaptive Gain Integrating Pixel Detector (AGIPD) and measure intensity correlations at different relative positions across the beam in a single measurement.

The AGIPD \cite{AGIPD1, AGIPD2, AGIPD3} is a novel detector system designed for the use at the European XFEL \cite{AltarelliTDR2007}.
It is aimed towards the demanding requirements of this machine for 2D imaging systems.
AGIPD is based on a hybrid pixel technology and operates most effectively in the energy range between 3 and 15~keV.
The current design goals are a dynamic range of more than 10$^4$ per pixel for 12.4~keV photons in the lowest gain, single photon sensitivity in the highest gain, and, importantly for XFEL applications and for our HBT experiment, operation at a frame rate of multiple MHz. %4.5~MHz frame rate.
Charges are stored in a memory bank inside each pixel and can be read out between the x-ray bursts, allowing to effectively capture and read out several thousand frames per second.

For our experiment, we used an AGIPD 0.4 assembly, which is a 16x16 pixel prototype of the AGIPD \cite{AGIPD4}, bump bonded to a silicon sensor of 320~$\mu$m thickness, giving a quantum efficiency of approximately 44\% at 14.4~keV photon energy.
The pixels are 200 $\times$ 200~$\mu$m$^2$ in size and feature the adaptive gain switching amplifier with 3 stages and 352 storage cells per pixel.
The prototype was read out via a chip testing box, which restricted the readout to less than 100 frames per second.
%The depletion voltage is approximately 50~V.
%To (over-)deplete the sensor, a bias voltage of 120~V was applied.
%{\it It is not clear for me how much details we need here}

\begin{figure}[t]
\includegraphics[width=1.\columnwidth]{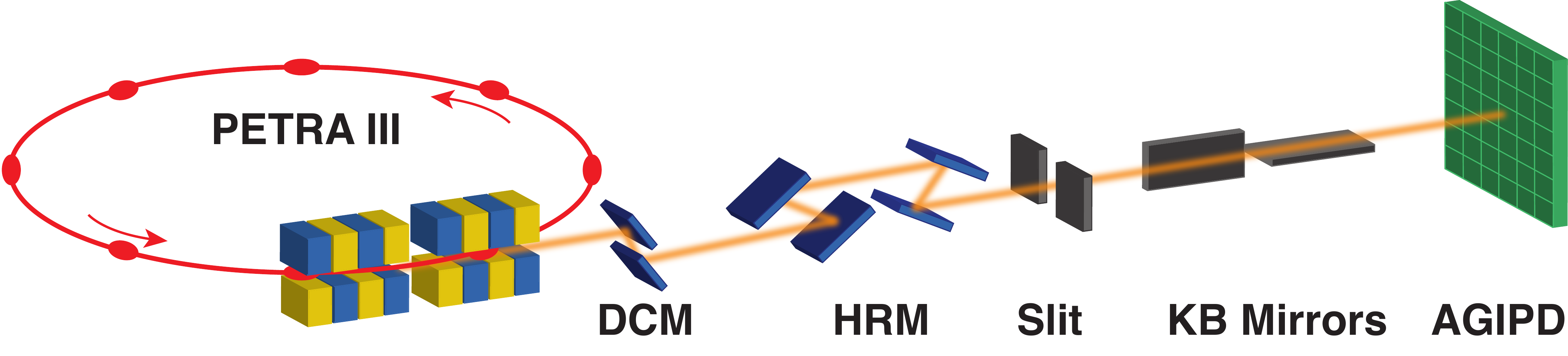}
\caption{
Experimental setup.
Synchrotron radiation is generated by the two five meter undulators and is transmitted through a double crystal monochromator (DCM), high resolution monochromator (HRM), slit system and set of KB mirrors.
Individual synchrotron radiation pulses were recorded with the AGIPD.
}
\label{fig:Sketch}
\end{figure}

The experiment was performed at the dynamics beamline P01 (Ref. \cite{P01Ref}) at the high-brilliance storage ring PETRA III, which is dedicated to inelastic X-ray scattering and nuclear resonant scattering.
The beamline layout is shown schematically in Figure \ref{fig:Sketch}.
Two five meter U32 undulators produced radiation at a photon energy of 14.4 keV.
The radiation was monochromatized with a double crystal monochromator (DCM) and high resolution monochromator (HRM) positioned at 48.5 m and 59.9 m downstream from the source, respectively.
The transmitted flux was about 10$^{10}$ photons/s at a bandwidth of 0.9 meV at 14.4 keV photon energy.
To reduce the number of transverse modes in the horizontal direction, a slit with a variable size was installed behind the monochromators at a distance of 61 m from the source.
%An optimum balance between a low number of transverse modes and high number of photons was found for a horizontal slit size of 200 $\mu$m.
The detector was positioned in the direct beam 94 m downstream from the source, and intensity profiles of individual synchrotron pulses were recorded.
%To effectively map out transverse coherence transverse coherence length has to be larger than  the pixel size.
To increase the vertical beam size as well as coherence length at the detector position, a Kirkpatrick-Baez (KB) system consisting of two mirrors
%coated by a multilayer in Bragg geometry
with a focal distance of 0.6 m was installed.
The distance from the KB system to the detector was 3 m, yielding a four-times magnification of the beam size and coherence length.
An increase of the transverse coherence length was required to determine the functional form of the correlation function with a higher resolution.
\footnote {We want to note here, that in the horizontal direction the transverse coherence length is significantly smaller and a larger magnification of the beam would be required to use the AGIPD for coherence measurements.}

\begin{figure}[b]
\includegraphics[width=1.\columnwidth]{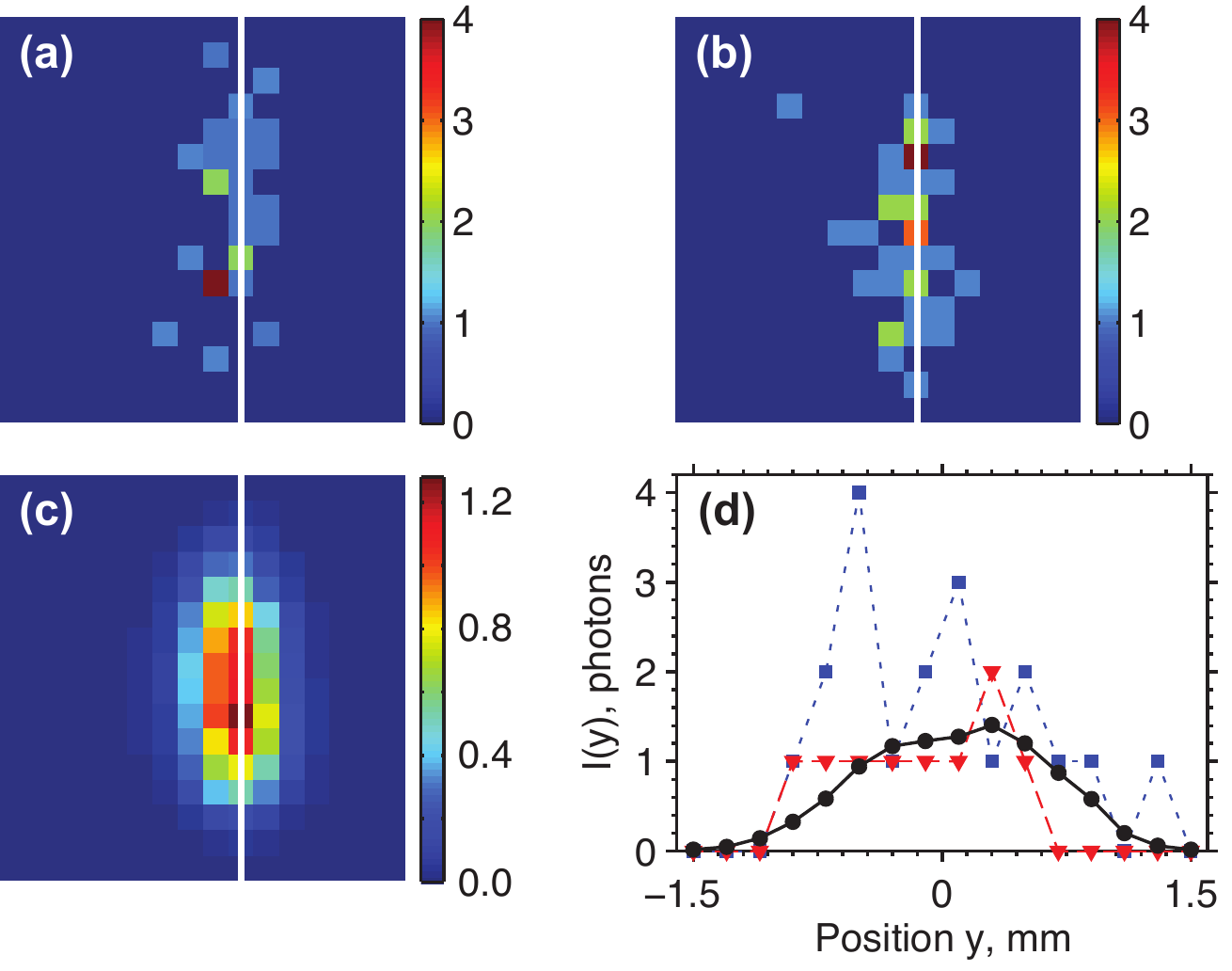}
\caption{Measured intensities.
(a,b) Typical intensity profiles of individual synchrotron radiation pulses recorded at the AGIPD.
The pixel size is 200 $\times$ 200 $\mu$m$^2$ and the detector size is 3.2 $\times$ 3.2 mm$^2$.
(c) Average intensity profile.
(d) Vertical line scans along white line in (a-c) through the typical single pulses (red triangles and dashed line (a), blue squares and dotted line (b)) and average intensity profile (black circles and solid line (c)).
}
\label{fig:Intensity}
\end{figure}

The detector was synchronized to the bunch repetition frequency of PETRA III (5.2 MHz), and about $3\cdot 10^5 $ intensity profiles of individual synchrotron pulses were recorded (see Fig.~\ref{fig:Intensity} (a,b)).
Figure \ref{fig:Intensity} (c) shows the average intensity profile.
The beam size at the detector position was $1.4\pm0.15$ (V) $\times$ $0.5\pm 0.1$ (H) mm$^2$ (FWHM) and covered the full detector in the vertical direction.
In the horizontal direction the size of the beam was defined by the slit size of 200 $\mu$m that provided an optimum balance between a low number of transverse modes and high number of registered photons.
Vertical line scans of the single-pulse 2D intensity profiles (see Fig.~\ref{fig:Intensity} (d)), were used for further analysis.
The pulses have significantly different profiles due to both the chaotic nature of the radiation and the low count rates.
At synchrotron sources the latter factor dominates the intensity fluctuations due to a small degeneracy parameter \cite{GluskinNIMA1992}.
%The pulses have significantly different profiles due to both the chaotic nature of the synchrotron radiation and the Poisson noise of the detection process.
%The Poisson noise was dominant in our experiment due to a small degeneracy parameter at synchrotron sources as opposed to recent experiments at an FEL \cite{SingerPRL2013}.

\begin{figure}[t]
\includegraphics[width=1.\columnwidth]{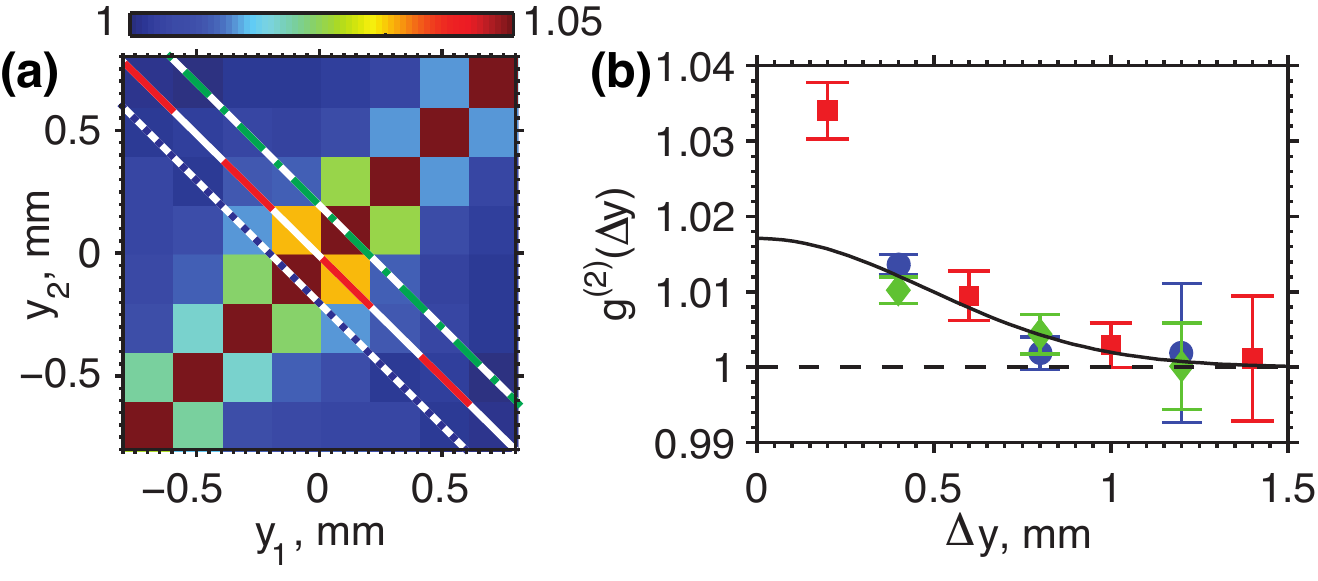}
\caption{Intensity correlation analysis.
(a) The normalized intensity correlation function $g^{(2)}(y_1,y_2)$.
(b) The normalized intensity correlation $g^{(2)}(\Delta y)$ as a function of $\Delta y$ for different cases: in the center of the beam along red dashed line in (a) (red squares), offset by a pixel to the left along blue dotted line in (a) (blue circles) and to the right along green dash dot line in (a) (green diamonds)
Fit by a Gaussian function (black solid line) provided transverse coherence length values of $l_c = 0.68\pm0.3$ mm (rms) (see Supplementary Material).
%The uncertainties were obtained by dividing the total ensemble of pulses into sub-ensembles each consisting of $5\cdot10^4$ pulses (see text for details).
}
\label{fig:Result}
\end{figure}

The result of the second order correlation function analysis (Eq.~\eqref{eq:2nd_corr}) along the vertical line shown in Fig.~\ref{fig:Intensity} is presented in Fig.~\ref{fig:Result}(a).
The normalized correlation function $g^{(2)}(y_1,y_2)$ shows the expected behaviour of maximum values for small separations between the pixels and a smooth fall-off for larger separations.
To determine the transverse coherence length and contrast we extract the intensity correlation function $g^{(2)}(\Delta y)$, $\Delta y=y_2-y_1$ from $g^{(2)}(y_1,y_2)$.
Figure \ref{fig:Result} (b) shows $g^{(2)}(\Delta y)$ for three different cases: in the center of the beam and offset by  200 $\mu$m (one pixel) to the left or to the right from the beam center
\footnote{
To estimate the statistical errors of the measurements, we divided the measured ensemble of pulses into eleven sub-ensembles, each consisting of $3\cdot10^4$ pulses, and used the spread of the resulting intensity correlation as an error measure.
%An error estimate was performed by two different ways.
The pulses were randomly permuted prior to the error analysis.
%In one case, pulses measured at similar times were put in the same ensembles, in the other case the pulses were randomly permuted (dashed and solid error bars in Figure \ref{fig:Result} (b), respectively).
The uncertainties obtained from the unpermuted ensemble are larger, which suggests drifts in the beamline during the measurement.
}.
Note that the values for both offsets are approximately equal, suggesting that the beam was quasi-homogeneous as expected from the theory \cite{VartanyantsNJP2010}.
Due to the low flux per pulse the values along the diagonal $y_1=y_2$ were dominated by the photon statistics and were not considered in further analysis (see Supplementary Material).
The normalized correlations between neighboring pixels $y_1 = y_2 \pm 0.2$ mm in Fig.~\ref{fig:Result} (a) show higher values than expected.
We attribute these high values to the parallax effect and discard these points from further analysis (see Supplementary Material).

The data shown in Fig.~\ref{fig:Result} (b) are well reproduced by a Gaussian function fit, $1 + \zeta\cdot\exp[(-\Delta y^2)/l_c^2]$ (see Eq.~\eqref{eq:chaotic}), which yields a transverse coherence length of $l_c = 0.68\pm0.3$ mm (rms) and a contrast value of $\zeta = 1.7\pm0.3$ \%.
The contrast is in good agreement with an estimate using PETRA III bunch parameters \cite{petra3}.
%the ratio $\tau_c/T$ (see eq. \eqref{eq:chaotic}), where $\tau_c=2\pi/\Delta\omega$ and $\Delta\omega$ is the frequency bandwidth \cite{G1985}.
From the energy bandwidth of 0.9 meV FWHM at 14.4 keV we find the coherence time of $\tau_c=2\pi/\Delta\omega \approx 4.6$ ps, which, together with a pulse duration of $96\pm3$ ps (FWHM) at normal operation conditions of PETRA III, yields a contrast of $\zeta \approx 4.8$ \%
if the horizontal transverse modes are neglected.
With these modes present the expected contrast value should be lower.

We have used the results of the transverse coherence measurements to determine the size of the synchrotron source.
Using the Gaussian Schell-model \cite{VartanyantsNJP2010} we found a source size of $7.8\pm3$ $\mu$m (see Supplementary Material).
This value is in excellent agreement with the photon source size $\sigma_y = 7.5$ $\mu$m estimated from the design electron beam parameters of the PETRA III storage ring \cite{petra3} (see also Supplementary Material).

In summary, we have demonstrated that the AGIPD can be used to measure intensity profiles of individual synchrotron pulses and to determine the transverse coherence properties of synchrotron radiation.
Using a prototype of this pixelated detector we were able to record the intensity correlation function at different relative spatial separations simultaneously.
A vertical photon source size of $7.8$ $\mu$m was determined from transverse coherence measurements.
This value agrees well with theoretical estimates based on the PETRA III storage ring parameters, which yield a value of $7.5$ $\mu$m.
We anticipate that this technique can be extended to hard x-ray free-electron lasers \cite{EmmaNatPhot2010,IshikawaNatPhot2012,AllariaNatPhot2012,SingerPRL2013}
and will provide a valuable diagnostic tool for next generation x-ray sources.
Due to the large frame rate, the final AGIPD detector will certainly be useful for a variety of experiments at high brilliance x-ray sources, for example, for a study of the dynamics of matter from the nano to microsecond time scales.

Part of this work was supported by BMBF Proposal 05K10CHG `'Coherent Diffraction Imaging and Scattering of Ultrashort Coherent Pulses with Matter`'
in the framework of the German-Russian collaboration `'Development and Use of Accelerator-Based Photon Sources`' and the Virtual Institute VH-VI-403 of the
Helmholtz Association.

%\bibliography{references}

%merlin.mbs aipnum4-1.bst 2010-07-25 4.21a (PWD, AO, DPC) hacked
%Control: key (0)
%Control: author (8) initials jnrlst
%Control: editor formatted (1) identically to author
%Control: production of article title (-1) disabled
%Control: page (0) single
%Control: year (1) truncated
%Control: production of eprint (0) enabled
%

\renewcommand{\theequation}{S\arabic{equation}}
\renewcommand{\thefigure}{S\arabic{figure}}
\section*{Supplementary Material}

\subsection*{Derivation of equation (2) for $\mathbf r_1=\mathbf r_2$}

For $\mathbf r_1=\mathbf r_2=\mathbf r$, equation (1) of the main text simplifies to
\begin{equation}
	g^{(2)}(\mathbf r,\mathbf r)=\frac{\avg{I(\mathbf r)^2}}{\avg{I(\mathbf r)}^2},
\label{eq:1}
\end{equation}
and can be written in terms of the variance $\textrm{Var}(I(\mathbf r))=\avg{I(\mathbf r)^2} - \avg{I(\mathbf r)}^2$
\begin{equation}
	g^{(2)}(\mathbf r,\mathbf r)=1 + \frac{\textrm{Var}(I(\mathbf r))}{\avg{I(\mathbf r)}^2}.
\label{eq:2}
\end{equation}
For chaotic light $I(\mathbf r)$ is distributed according to the negative binomial distribution and the variance is given by \cite{G1985}
\begin{equation}
	\textrm{Var}(I(\mathbf r)) = \frac{\bar{N}^2}{M} + \bar{N} ,
\label{eq:3}
\end{equation}
where $\bar{N} =  \avg{I(\mathbf r)}$ is the average number of photons and $M$ is the number of modes.
Substituting equation \eqref{eq:3} into equation \eqref{eq:2} yields
\begin{equation}
	g^{(2)}(\mathbf r,\mathbf r)=1 + \frac{1}{M}+ \frac{1}{\bar{N}} .
\label{eq:4}
\end{equation}
Equation (2) of the main text for identical pixels follows immediately using $|\gamma(\mathbf r,\mathbf r)|=1$.

\newpage
\subsection*{Average intensity and intensity correlation measurement uncertainty}

\begin{figure}[b]
\includegraphics[width=1.\columnwidth]{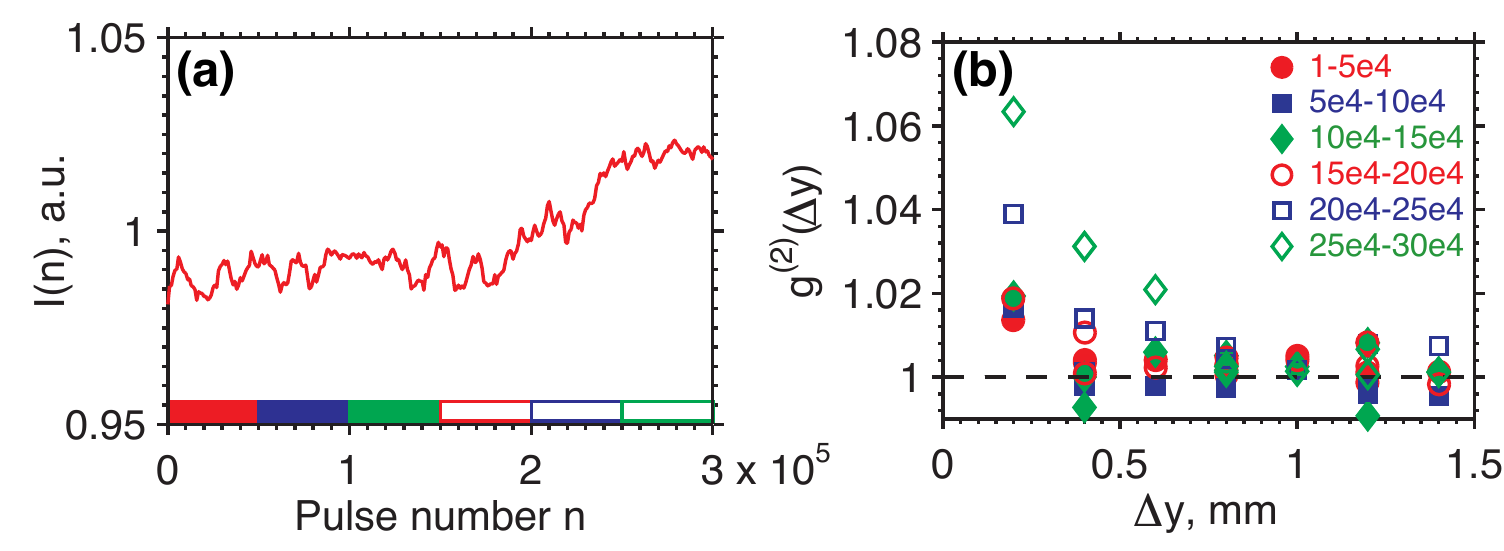}
\caption{
(a) Averaged intensity per pulse taken along the line scan shown by the white line in Figure 2 (a) of the main text.
The values have been smoothed by a rectangular function $10^4$ pulses wide.
(b) The correlation function as shown in Figure 3 (b) of the main text for six different sub ensembles of $5 \cdot 10^4$ pulses each.
The respective range of the analyzed pulses is indicated in (a).
}
\label{fig:S1}
\end{figure}

The average intensity as a function of the pulse number observed during the experiment is presented in Figure \ref{fig:S1} (a).
Small fluctuations (smaller than 1\%) were observed during most of the experiment.
However, there is also a distinct increase in intensity of about 3\% for pulse numbers higher than $2\cdot 10^5$.
The intensity correlation function calculated from different parts of the total number of pulses is shown in Figure \ref{fig:S1} (b).
The values of the correlation function appear to be slightly higher in the higher intensity region observed in Figure \ref{fig:S1} (a).
The result presented in the main text is the average over the whole ensemble of pulses.
The measurements hence demonstrate that, while a large number of pulses reduces the statistical errors, there can be drifts in the beamline components during the resulting large measurement times.

\newpage
\subsection*{Enhanced correlation between neighboring pixels}

If we enumerate all the pixels along a stripe, then (linear) crosstalk between neighboring pixels gives the relation
\begin{equation}
	J_i = k_{i-1,i} I_{i-1} + I_i + k_{i+1,i} I_{i+1}
\end{equation}
between the recorded intensity $J_i$ of pixel $i$ and the incident intensities $I_i$.
The value $k_{i,j}$ gives the amount of crosstalk between pixels $i$ and $j$.

Let us assume that the crosstalk is homogeneous, $k_{i,j}=k$, and small, $k\ll1$.
Dropping all terms of order $k^2$, we can calculate the correlation functions as
\begin{eqnarray}
\label{eq:numerator}
	\avg{J_1 J_2} &\approx& \avg{I_1I_2} + k \Bigl( \avg{I_0I_2} + \avg{I_1I_3} + \avg{I_1^2} + \avg{I_2^2} \Bigr) \\
\label{eq:denominator}
	\frac{1}{\avg{J_1}\avg{J_2}} &\approx& \frac{1}{\avg{I_1}\avg{I_2}} \Bigl(1 - k\eta\Bigr),
\end{eqnarray}
with $\eta = \avg{I_0}/\avg{I_1} + \avg{I_3}/\avg{I_2} + \avg{I_1}/\avg{I_2} + \avg{I_2}/\avg{I_1}$, and where we used a lowest-order Taylor-expansion in equation~\eqref{eq:denominator}.

We then insert equations \eqref{eq:numerator},\eqref{eq:denominator} into equation (1) of the main text and drop all terms of order $k^2$ or $k\zeta$.
This gives the final result to first order in $k$, $\zeta$ as
\begin{equation}
g^{(2)}_{12} =
1+\zeta\cdot|\gamma_{12}|^2 + k \left(\frac{1}{\avg{I_1} } + \frac{1}{\avg{I_2}} \right) .
\label{eq:8}
\end{equation}

In the presence of crosstalk, the apparent correlation function for neighboring pixels is higher than the true correlation.
This effect is especially important for small photon numbers $\avg{I_i}$.
We therefore attribute the high value for the correlation function at neighboring pixels in the main text to such crosstalk.

\begin{figure}[h]
\includegraphics[width = 0.3\textwidth]{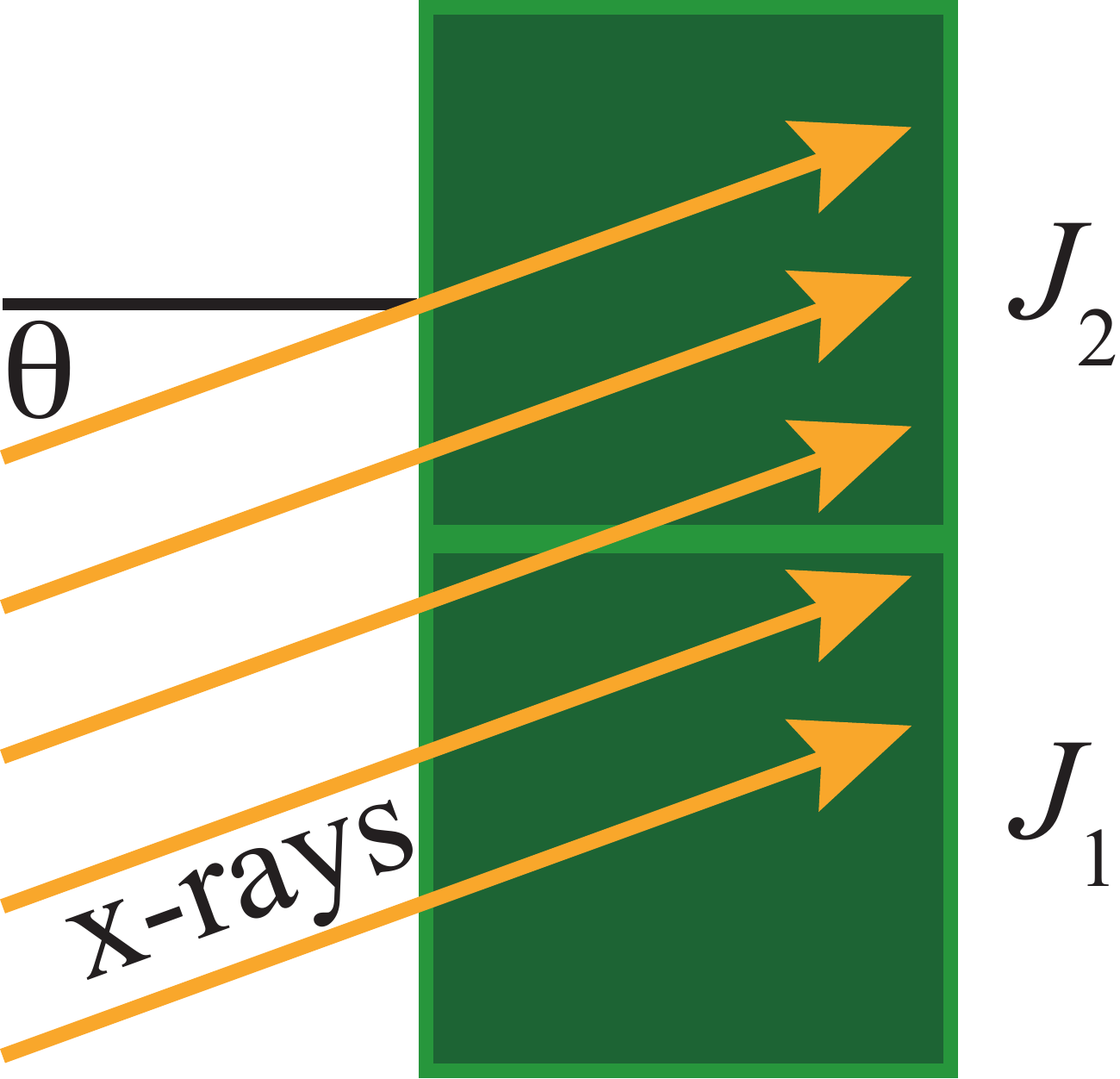}
\caption{An illustration of the parallax effect.}
\label{fig:fS3}
\end{figure}
A similar effect can be observed if the detector is slightly misaligned with respect to the incident beam (see Figure \ref{fig:fS3}), the so called parallax effect.
In the case of the parallax effect the increased correlation between neighboring pixels acts only in one direction, the direction of the inclination angle.
The recorded intensities $J_i$ are then given by
\begin{equation}
\begin{split}
J_1 &= I_1\\
J_2 &=I_2 + kI_1,
\end{split}
\end{equation}
where $I_i$ are the incident intensities and $k$ describes the increased correlation between neighboring pixels.
After calculations similar to \eqref{eq:numerator}-\eqref{eq:8} we find in the case of the parallax effect
\begin{equation}
g_{12}^{(2)} = 1 + \zeta\cdot |\gamma_{12}|^2+k\frac{1}{\avg{I_1}}
\label{paral}
\end{equation}

We can estimate the magnitude of $k$ from equation \eqref{paral} and the result presented in the main text.
The intensity $I_1\approx1.2$ (see Figure 2(d) of the main text) and the offset
$k/\avg{I_1}\approx0.015$ (compare the value of the measured correlation $g_{12}^{(2)}$ and the fit in Figure 3(b) of the main text), which yields $k\approx0.01$. From the pixel properties and pixel separation we estimate an inclination angle of $\theta = 1.4$ degree.
The detector was aligned for the measurement without the KB mirror system.
The KB optic at P01 reflects the beam at a Bragg angle of 0.8 degree and the exit angle is two Bragg angles.
This means the detector was $\theta=1.6$ degrees off in the measurements presented in the main text.
This value concords well with the estimate from the experiment.

\newpage
\subsection*{Uncertainty of the transverse coherence length}

\begin{figure}[b]
\includegraphics[width=1.\columnwidth]{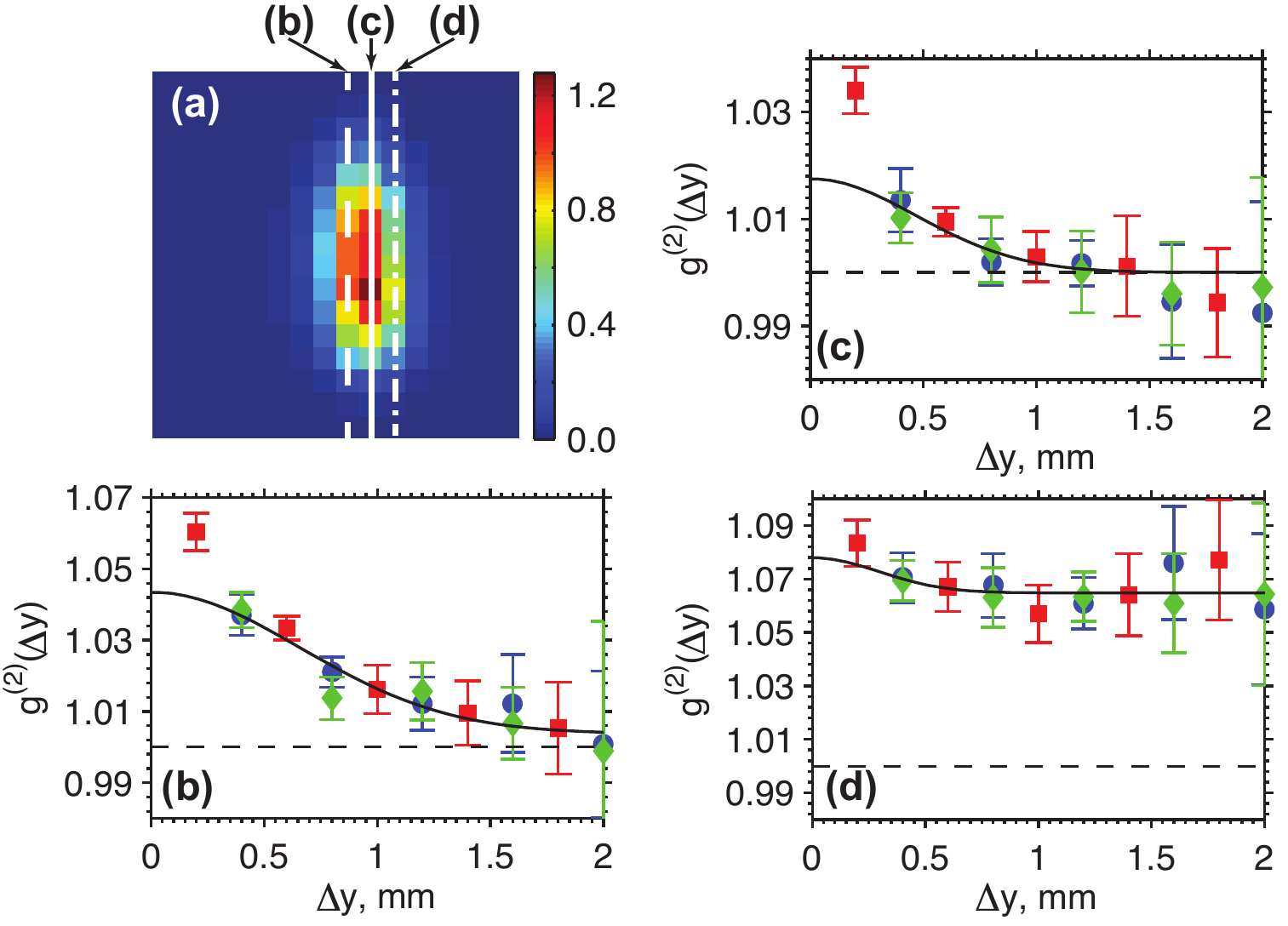}
\caption{
Analysis of different regions of measured intensities.
(a) The average intensity profile as shown in Figure 2(c) of the main text.
Correlations calculated from single pulse intensity line profiles along the solid (a), dashed (b), and dash-dotted (c) line in (a).
}
\label{fig:S2}
\end{figure}

To estimate the uncertainty of the transverse coherence length in the main text we have done the analysis presented in Figure 3 of the main text for different regions of the detector.
In particular intensity profiles along three different lines were calculated: the line with the maximum intensity (center, also shown in the main text), and the two lines one pixel distant (see Figure \ref{fig:S2}).
The resulting coherence lengths are $0.68 \pm 0.1$ mm for the central line (Figure \ref{fig:S2} (c)), and $0.93\pm0.2$ mm and $0.4\pm0.4$ mm for the non-centered lines (Figure \ref{fig:S2} (b) and (d), respectively).
It should be noted that the uncertainty of the coherence length for the central line is considerably smaller than the variance from the different lines.
To account for the variation between different lines we have extended the uncertainties of the result from the main text to yield a value of $0.68\pm0.3$.
%The intensity correlation function calculated from the horizontally averaged intensity profiles yields a coherence length of $0.5\pm0.1$ mm.

As in the main text, we have not included the correlation between neighboring pixels in the fit.
This value is consistently larger than the fit and we attribute it to a parallax effect.
If we include this point to the fit, the coherence length decreases by about 30\%.

The intensity correlation function obtained from the off center pixel columns shows a constant background (see Figure \ref{fig:S2} (b,d)), which is particularly large in Figure \ref{fig:S2} (d).
A possible explanation of this behavior is the lower intensity in the off center regions of the detector.

\newpage
\subsection*{Determination of the source size from electron beam parameters and the transverse coherence measurements.}

The photon source size and divergence are given by $\sigma = \sqrt{\sigma_e^2+\sigma_r^2}$ and $\sigma' = \sqrt{\sigma_e'^2+\sigma_r'^2}$, respectively.
They depend on the electron beam size $\sigma_e=\sqrt{\varepsilon\beta}$ and divergence $\sigma_e'=\sqrt{\varepsilon/\beta}$ as well as the parameters of radiation emitted by a single electron: size $\sigma_r=\sqrt{2\lambda L_u}/4\pi$ and divergence $\sigma_r' = \sqrt{\lambda/2Lu}$ \cite{VartanyantsNJP2010}.
Here $\varepsilon$ is the electron beam emittance of the storage ring, $\beta$ is the betatron function value in the center of the undulator, $L_u$ is the length of the undulator, and $\lambda$ is the wavelength of x-ray radiation. We used the following parameters of the source:
electron beam emittance $\varepsilon = 0.01$ nmrad, betatron function $\beta = 4.5$ m, and undulator length $L_u=10$ m.

To calculate the source size from the coherence measurements we have assumed that the high resolution monochromator does not affect the transverse coherence length in the vertical direction since it does not modify the divergence of the beam \cite{SouvorovJPD1999}.
The KB mirrors were considered as a perfect imaging system \cite{SingerJSR2014} that magnifies the transverse coherence length by a factor of four.
%The measured beam size is smaller than expected from simulations.
%We attribute this disagreement to a finite acceptance width of the KB system.
In this approximation the source size was calculated using equation (29) from Ref.  \cite{VartanyantsNJP2010}
\begin{equation}
\sigma = \frac{z}{2kl_c}\sqrt{4+q^2}
\label{eq:5}
\end{equation}
where $q=l_c/\Sigma$, $l_c$ and $\Sigma$ are  the  transverse coherence length and beam size measured at the detector.
The beam size measured with the KB mirror system was smaller than expected from the geometry.
We attribute this disagreement to a finite acceptance width of the KB system and assume that the finite acceptance of the mirror does not affect transverse coherence \cite{SingerJSR2014}.
In equation \eqref{eq:5} the beam size $\Sigma$ was obtained from the divergence of the source $\sigma_y' = 5.5\pm1$ $\mu$rad, which was determined from a beam size of $0.52\pm0.05$ mm measured at the detector without focusing optics.
This value is also in good agreement with the divergence $\sigma_y' = 6 ~\mu$rad expected from the electron beam parameters.

\end{document}